\documentclass[12pt]{article}
{\begin{Sbox}\begin{minipage}}%
		{\end{minipage}\end{Sbox}\fbox{\TheSbox}}%
\usepackage[psamsfonts]{amssymb}
\usepackage{amsmath, amsthm, anysize, enumerate, color, url}
\allowdisplaybreaks[4]
\usepackage{graphicx}
\usepackage{epstopdf}
\usepackage{epsfig}
\usepackage{subfigure}
\usepackage{multirow}
\usepackage[ruled,linesnumbered]{algorithm2e}
\usepackage{tikz}
\usepackage{array}

\usepackage[sectionbib]{natbib}

\usepackage{threeparttable}

\usepackage[colorlinks, breaklinks,
linkcolor=red,
citecolor=blue
]{hyperref}
\usepackage{breakurl}
\usepackage{bm}

\newtheorem{theorem}{Theorem} 

\newtheorem{lemma}{Lemma} 
\newtheorem{proposition}{Proposition} 

\newcommand{\balign}{\begin{align}}
	\newcommand{\ealign}{\end{align}}
\newcommand{\beq}{\begin{equation}}
	\newcommand{\eeq}{\end{equation}}

\theoremstyle{definition}
\newtheorem{remark}{Remark}  

\newcommand{\E}{\mathbb{E}}

\renewcommand{\P}{\mathbb{P}}

\newcommand{\diag}{\mbox{diag}}

\usepackage{setspace}
\allowdisplaybreaks
\usepackage{xr}

\begin{document}
\title{\bf Subgraph counting estimation for the $\beta$-model in sparse networks \footnote{Author names are sorted alphabetically.}}

\author{Qunqiang Feng\thanks{Department of Statistics and Finance, University of Science and Technology of China, Hefei, 230000, China.
		\texttt{Email:} fengqq@ustc.edu.cn}
	\hspace{4mm} \and
	Jiashun Jin\thanks{Department of Statistics, Carnegie Mellon University, Pittsburgh, PA 15213, USA.
		\texttt{Email:} jiashun@stat.cmu.edu}
	\hspace{4mm} \and
	Yaru Tian \thanks{School of Statistics \& Data Science, Southeast University, Nanjing, 211189, China.
		\texttt{Email:} }
	\hspace{4mm} \and
	Ting Yan  \thanks{Department of Statistics, Central China Normal University, Wuhan, 430079, China.
		\texttt{Email:} tingyanty@mail.ccnu.edu.cn.}
}
\maketitle
		
\begin{abstract}
	\begin{spacing}{1.2}
	The $\beta$-model is popular for characterizing the commonly observed degree heterogeneity phenomenon in real-world networks.
	In this study, we develop a cycle counting approach to estimate $n$ node-specific parameters in the $\beta$-model for moderate or extremely sparse networks.
	Our proposed estimators, called \emph{Cycle Counting  Ratio (CCR) Estimator},
	are based on the log-ratios of two network cycle counting statistics with explicit expressions and therefore easy to compute.
	We focus on conditions to guarantee statistical properties of the single estimator for each node.
	Under the very weak conditions that $\max_t \theta_t \to 0$ and $\theta_t \|\theta\|_1 \to \infty$, we show that the CCR estimator is consistent and achieves the minimax rate in terms of the mean squared error, which is the squared signal-to-noise ratio for $\hat{\beta}_t$ up to a constant factor. 
	Here, $\hat{\beta}_t$ is the CCR estimator of the node-specific parameter $\beta_t$, $\theta_t = \exp(\beta_t)$
	and $\theta=(\theta_1, \ldots, \theta_n)$.
	Even if the whole network density is close to the  Erd\H{o}s-R\'{e}nyi lower bound $\log n/n$, the CCR estimator for the single parameter $\beta_t$ is still consistent as long as $\theta_t \|\theta\|_1 \to \infty$.
	To the best of our knowledge, this is the first time to derive the minimax rate and consistency result under such weak conditions.
	Under a slight stronger condition, we further establish its uniform consistency and asymptotic normality, whose asymptotic variance is $\theta_t \|\theta\|_1$.
	Numerical studies and an application to a sparse network data set demonstrate our theoretical findings.
	\end{spacing}
	
	\begin{spacing}{1.4}
	\end{spacing}
\end{abstract}

	\section{Introduction}

	The $\beta$-model \citep{Chatterjee2011random} is an exponential random graph model with the degree sequence as the
	natural sufficient statistic. Specifically, it assumes that a random element $A_{ij}\in\{0,1\}$
	of the adjacency matrix $A$ of an undirected graph
	is independently distributed as a Bernoulli random variable with the link probability:
	\begin{equation}
		\label{model}
		\P(A_{ij}=1)=\frac{e^{\beta_i+\beta_j}}{1+e^{\beta_i+\beta_j}},
	\end{equation}
	where the parameter $\beta_i$ measures the strength of node $i$ to participate in network connection and
	is generally referred to as the degree heterogeneity parameter (degree parameter for short).

	The $\beta$-model can be viewed as an undirected version of an earlier $p_1$ model
	for directed graphs \citep{Holland1981exponential}.
	It has been widely applied to model the degree heterogeneity phenomenon in real-world networks \citep[e.g.][]{Park:Newman:2004,Blitzstein:Diaconis:2011,Chen:2020},
	which describes variations in the number of edges amongst nodes.
	In addition, it serves as a null model for hypothesis testing [\cite{Holland1981exponential,Fienberg:Wasserman:1981,Zhang:Chen:2013}] and can be
	used to reconstruct networks and make statistical inferences when only the degree sequence is available,
	owing to privacy considerations
	[\cite{Helleringer:Kohler:2007,shao2023l2}]. It can be also used in a preliminary
	analysis to choose suitable statistics for network configurations
	[\cite{Robins.et.al.2009}].
	
	Since the number of parameters in the $\beta$-model is the same as the number of nodes, statistical inference is non-standard
	and attracts great interests from many statisticians
	\citep[e.g.][]{Chatterjee2011random,Yan2013clt,Rinaldo2013,Karwa:Slakovic:2016,yan2025likelihood},
	where the conditions for the existence of the maximum likelihood estimator (MLE), its consistency and asymptotic normality are derived.
	However, these works focus on relatively dense networks, where the network density, defined as the ratio of the expected number of observed edges
	to the total number of all possible edges, is either a constant or goes to zero with a very slow rate.
	In moderate or extremely sparse networks,  the maximum likelihood estimation may suffer from the problem of the non-existence, as demonstrated in \cite{Yan2013clt}.

	The penalized likelihood methods have been proposed to draw inference, including
	the $\ell_0$-penalty in \cite{Chen:2020}, the $\ell_1$-penalty in \cite{stein2020sparse} and the $\ell_2$-penalty in \cite{shao2023l2}.
	\cite{Chen:2020} required a strong condition on the $\beta$-parameter structure, where most of parameters are essentially equal and
	network densities are larger than $O(n^{-1/2})$.
	\cite{stein2020sparse} required that network densities are larger than $O(n^{-1/6})$.
	The regularized MLE in \cite{shao2023l2} could handle a very sparse network with density being close to $O(\log n/n)$, but it needs to
	assume that almost all parameters are equal.
	In a general model parameter configuration, the regularized MLE
	incurs the bias problem.
	We shall elaborate these related works in the next subsection.
	As a result, how to effectively make accurate estimation in the $\beta$-model under weak assumptions on the model parameters in large and sparse networks,
	is still an open question.

	In this study, we develop the subgraph counting approach with a cycle format to estimate the unknown parameters in the $\beta$-model.
	Our proposed estimator, called \emph{Cycle Counting Ratio (CCR) Estimator}, is based on the log-ratio of two sums for the $m$-cycle counts
	in terms of
	\[
	\sum_{i_1,\ldots,i_m} A_{i_1, i_2}^{b_1}\cdots A_{i_{m-1}, i_m}^{b_{m-1}}A_{i_{m-1}, i_m}^{b_m}A_{i_{m}, i_1}^{b_{m+1}},
	\]
	where $b_i\in\{0,1\}$. 
	The central idea behind it is that the unknown parameters
	can be represented as the logarithm of ratios of the probabilities of observing two different subgraphs.
	It is remarkable that the cycle counting approach has been used in \cite{JinGC2018,jin2019optimal} for community detection problems and \cite{feng2026optimal} for estimating the reciprocity parameter in the $p_1$ model.
	Although there are cases for $m> 3$ cycle counts (see Proposition \ref{propo:ab}), where $\beta_i$ can be represented as
	the logarithm of ratios of the probabilities of observing two different $m$-cycles, we consider $m=3$ here.
	This is because theoretical analysis is simpler in case of cycles of length $3$ than those cycles of larger lengths. In addition,
	the computational cost is also lower.
	Another reason is that there are no significant improvements in the estimation accuracy in a comparative simulation study for $m> 3$.
	Except for the isomorphic cycles, there are only one pair of $3$-cycles, in which the parameters can be expressed as
	the logarithm of ratios of their probabilities.
	Our proposed estimators	have explicit expressions,
	unlike the MLE or its penalized version that has to resort to iterative algorithms because its explicit solution is not
	possible.
	Therefore, our estimation can be scaled to very large networks with millions of nodes,
	in which the $\ell_0$ or $\ell_1$ penalized likelihoods
	face computational challenge [\cite{Chen:2020,stein2020sparse}].

	Our contribution is threefold.
	First,
	under the very weak conditions $\max_t \theta_t \to 0$ and $\theta_t \|\theta\|_1 \to \infty$, we show that
	the CCR estimator $\hat{\beta}_t$ of the parameter $\beta_t$ for node $t$ achieves the minimax rate in terms of the mean squared error,
	which is the squared signal-to-noise ratio for $\hat{\beta}_t$ up to a constant factor.
	Here, $\theta_t = \exp(\beta_t)$
	and $\theta=(\theta_1, \ldots, \theta_n)$.
	Second, we establish  consistency of $\hat{\beta}_t$ under the same conditions.
	Even if the whole network density is close to the  Erd\H{o}s-R\'{e}nyi lower bound $\log n/n$ \citep{erd6s1960evolution},
	the CCR estimator for the single parameter $\beta_t$ is still consistent as long as $\theta_t \|\theta\|_1 \to \infty$,
	To the best of our knowledge, this is the first time to derive the minimax rate and consistency result
	under such weak conditions. We further establish the uniform consistency of all estimators under additional conditions.
	Third,	under a slight stronger condition, up to a logarithmic factor, we establish its asymptotic normality, whose asymptotic variance is $\theta_t \|\theta\|_1$.
	Numerical studies and a real data application illustrate our theoretical findings.

	\subsection{Literature review}
	
	We give a comprehensive literature review on the $\beta$-model.
	As mentioned before, asymptotic inference in the $\beta$-model is nonstandard due to the increasing number of parameters.
	Under the assumption $\max_i |\beta_i| \le C$ with a positive constant $C$, when the number of nodes goes to infinity,
	\cite{Chatterjee2011random} established the uniform consistency of the MLE via obtaining the geometrically fast convergence rate
	of a fixed-point iterative algorithm for solving the MLE.
	Under the condition $\max_i |\beta_i| =o(\log (\log n))$,
	\cite{Yan2013clt} proved its asymptotical normality by approximating the Fisher information matrix.
	\cite{Rinaldo2013} derived the necessary and sufficient conditions of its existence of the MLE, but they are hard to verify.
	\cite{Karwa:Slakovic:2016} further gave easily checked conditions for judging the existence of the MLE.
	Under the large sample framework that the number of edges are observed many times and the number of nodes is fixed, \cite{wahlstrom2017beta} established the Cram\'{e}r-Rao bounds
	and presented maximum likelihood estimators to test  significance of parameters in the $\beta$-model.

	In order to  handle sparse networks,
	\cite{Chen:2020} assumed a $\beta$-sparsity structure with a so-called active set $S$, where
	the $\beta$-parameters are divided into two parts: $\beta_i=-(\gamma/2) \log n + o(\log n)$, if $i\in S$;
	$\beta_i=( \alpha - \gamma/2)\log n + o(\log n)$, if $i\notin S$.
	Here,  $\alpha$ and $\gamma$ are known constants.
	Then, they introduced the $L_0$-penalty likelihood method to estimate the unknown parameters.
	However, the theoretical properties of the penalty likelihood estimator, including consistency and asymptotic distribution, are not investigated in their paper.
	A subsequent work by \cite{wang2025group} relaxes the homogeneity assumption of degree parameters in the active set
	by assuming that these core nodes can be divided into several groups with different orders of magnitude of parameters.
	When the groups are known, they provide consistent and asymptotically normal moment estimators of the parameters.
	However, the latent groups in real-world networks are unknown.
	In this case, how to derive consistent estimators for  unknown parameters is not addressed in their paper.
	\citet{stein2020sparse} proposed an $L_1$ regularized method to estimate an adjusted $\beta$-model.
	As pointed by \cite{shao2023l2}, the network density in \citet{Chen:2020} and \citet{stein2020sparse} are required to be larger than $O(n^{-1/2})$ or $O(n^{-1/6})$, respectively.
	\cite{shao2023l2} propose an $L_2$ regularized method
	and can handle sparser networks whose network density is  close to $\log n/n$,
	but requires a strong condition that all parameter are hardly the same.
	In contrast to these existing literature,
	we do not require the structured parameter assumption and establish consistency and asymptotically normal distribution of
	the 3-cycle-ratio estimator under the weakest condition on the network density.

	We now briefly review privacy issues related to the $\beta$-model.
	Under the assumption that all parameters are bounded above by a constant, \cite{Karwa:Slakovic:2016}
	used the output perturbation algorithm to release the degree sequence and
	constructed private estimators in the $\beta$-model that are consistent and asymptotic normal distributions when the number of nodes goes to infinity.
	\cite{chang2024edge} further designed an input perturbation algorithm to release network data
	and proposed a moment estimator for the unknown $\beta$-parameters.
	They show that it is consistent and asymptotic normality in dense networks and it is consistent in the sparse $\beta$-model \citep{Chen:2020}.
	It is remarkable that the estimator proposed by \cite{chang2024edge}  is the same as ours in case of no privacy protection.
	However, they focus on dense networks and privacy problems, which is different from ours.
	Here, we study sparse networks and weak conditions on parameter estimation.
	
	We conclude this section by mentioning some generalized $\beta$-models.
	\cite{Hillar:Wibisono:2013} generalized the $\beta$-model to weighted networks and obtained
	consistency under the same condition as in \cite{Chatterjee2011random}; \cite{Yan:Zhao:Qin:2015}
	derived the asymptotically normal distribution of the MLE.
	Further, \cite{Yan:Qin:Wang:2015} established a unified theoretical framework for these models.
	\cite{Yan:Leng:Zhu:2016}
	established consistency and asymptotical normality of the MLE 
	in the $p_0$ model for directed networks, which
	is a special case of the $p_1$ model without the reciprocity parameter.
	Other node-parameter models include the Chung-Lu model [\cite{Chung:Lu:2002}] with the expected degrees as the parameters,
	null models [\cite{Perry:Wolfe:2012}] and the hypergraph $\beta$-model [\cite{nandy2024degree}].
	\cite{Du2023sjs} extended the static $\beta$-model to dynamic directed networks.
	\cite{jiang2025two} introduced an autoregression network model with
	static and dynamic nodes' heterogeneity parameters.
	In addition, the covariates of nodes were also  incorporated into the $\beta$-model \cite[e.g.][]{wahlstrom2017beta,Graham2017}.

	The rest of the paper is organized as follows. Section \ref{section-estimation} introduces the estimation method. Section \ref{sec-theoretical} presents
	asymptotic optimality, consistency and asymptotic normality of the CCR estimator.
	Section \ref{section-simulation} presents simulation results. Section \ref{section-discussion} discusses our findings and concludes the paper.
	All the technical proofs are relegated into supplementary Material.

	\section{Estimation methods}
	\label{section-estimation}
	
	Consider an undirected graph $G_n$ with $n$ nodes, labelled as $[n]:=\{1,2,\cdots,n\}$.
	Let $A=(A_{ij})_{n\times n}$ be its adjacency matrix, where $A_{ij}\in \{0,1\}$ with $A_{ij}=1$ denoting an edge between nodes $i$ and $j$ and
	$A_{ij}=0$ denoting no edge. For convenience, set $A_{ii}=0$ over $i=1, \ldots, n$.
	The $\beta$-model assumes that
	$\{A_{ij},1\leq i<j\leq n\}$ are independent bernoulli random variables with link probabilities specified in \eqref{model}.

	We use the cycle counting approach to estimate the unknown parameters.
	We quote the notion of cycles from  \citet{JinGC2018,jin2019optimal,feng2026optimal}.
	For $s \ge 3$, a \emph{$s$-cycle} is a subgraph of $G_n$ consisting of $s$ distinct nodes $i_1, i_2, \ldots, i_s$
	and the set of edge types $\{ A_{i_1, i_2}, A_{i_2,i_3}, \ldots, A_{i_{s-1},i_{s}}, A_{i_s, i_1}\}$.
	We call an $s$-cycle of Type-$a$ if the edge types are $a_1, \ldots, a_s$, where $a_i \in \{0,1\}$.
	As an illustrating example, the $4$-cycle of Type-$a=(1,0,1,0)$ with nodes $i_1$, $i_2$, $i_3$ and $i_4$ means that
	there are edges for the pairs $(i_1,i_2)$ and $(i_3, i_4)$, and no edges between $(i_2,i_3)$ and $(i_4,i_1)$.

	Let $S_m = \{a = (a_1, a_2, \ldots , a_{m+1}): a_i\in \{0, 1\}, i=1,\ldots, m+1\}$.
	For a fixed $m \geq 2$ and $a \in S_m$, we define a cycle counting statistic for node $t$:
	\[
	G_{t,m}(a) = \sum_{ (i_1, i_2, \ldots, i_m) \in \mathcal{F}_m  } A_{t,i_1}^{a_1}A_{i_1,i_2}^{a_2} \cdots A_{i_{m-1}, i_{m}}^{a_{m}} A_{i_{m}, t}^{a_{m+1}},
	\]
	where $\mathcal{F}_{t,m}=\{(i_1, i_2, \ldots,i_m): i_1, i_2, \ldots, i_m \mbox{~are distinct and~} i_j\in [n]\backslash \{t\} \mbox{~for~} j=1,\ldots, m \}$.
	Let $g_{t,m}(a)= \E [ G_{t,m}(a)]$.
	Because $A_{t,i_1}, A_{i_1,i_2}, \cdots,  A_{i_m, t}$ are independent when $(i_1, i_2, \ldots, i_m)\in \mathcal{F}_{t,m}$, we have
	\begin{eqnarray*}
		g_{t,m}(a) & = &   \sum_{ (i_1, i_2, \ldots, i_m) \in \mathcal{F}_m }  \frac{1}{ (1+e^{\beta_t+\beta_{i_1}})(1+e^{\beta_{i_1}+\beta_{i_2}})\cdots
			(1+e^{\beta_{i_{m-1}}+\beta_{i_m}}) (1+e^{\beta_{i_m}+\beta_t})}
		\\
		&&\quad \cdot \exp\big\{ \underbrace{ \beta_t(a_1+a_{m+1}) +  \beta_{i_1} (a_1 + a_2) + \cdots + \beta_{i_{m-1}}(a_{m-1} + a_{m} )+ \beta_{i_m}(a_{m} + a_{m+1} ) }_{ f(a, \beta_{i_1}, \ldots, \beta_{i_m}) } \big\}.
	\end{eqnarray*}
	Observe that the factors before the exponential part in each term in the above big sum is the same for any $a\in S_m$.
	For $a, b\in S_m$, if
	\[
	f(a, \beta_{i_1}, \ldots, \beta_{i_m}) - f(b, \beta_{i_1}, \ldots, \beta_{i_m}) = c \beta_t,
	\]
	then  $g_{t,m}(a) = \exp( c\beta_t ) g_{t,m}(b)$ with $c\neq 0$.
	Formally, we have the following proposition, whose proof is easy and omitted.

	\begin{proposition}
		\label{propo:ab}
		For $a, b \in S_m$, if
		\begin{equation}\label{critical}
			a_{i-1} + a_i = b_{i-1} + b_i, ~~i=2,\ldots,m, ~~\mbox{and}~~ a_1 + a_{m+1} \neq b_1 + b_{m+1},
		\end{equation}
		then
		\[
		\frac{ g_{t,m}(a) }{ g_{t,m}(b) } = e^{ (a_1 + a_{m+1} - b_1 - b_{m+1}) \beta_t }.
		\]
	\end{proposition}

	\begin{figure}
		\centering
		\includegraphics[height=1in,width=4.55in, angle=0]{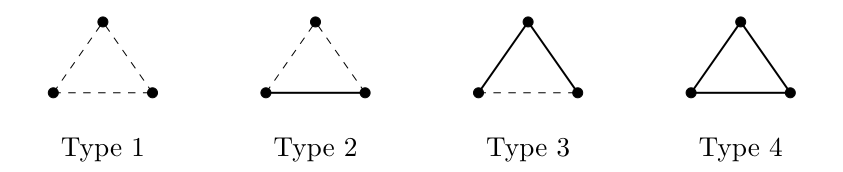}
		\caption{All non-isomorphic cycles with $3$ nodes.}
		\label{fig-types}
	\end{figure}
	
	This motivates us to estimate $\beta_t$ by $c^{-1} \log \big\{ G_{t,m}(a)/G_{t,m}(b) \big\}$.
	We need to find a pair $(a,b)\in S_m$ satisfying \eqref{critical}.
	Consider the case $m=2$, which corresponds to a $3$-cycle.
	There are a total number of $4$ non-isomorphic cycles with three nodes in Figure \ref{fig-types}.
	For $a=(0,0,0)$ or $a=(1,1,1)$, there are no such $b\neq a$ satisfying \eqref{critical}.
	In fact, only the pair $(a,b)$ or $(b,a)$ satisfies \eqref{critical}, where $a=(1,0,1)$ and $b=(0,1,0)$.
	Because it is equivalent to estimate $\beta_t$ by using $(a,b)$ or $(b,a)$, we only consider $(a,b)$.
	As a result, we propose to estimate $\beta_t$ for $t=1,\ldots, n$ by
	\begin{equation}\label{hat-beta-star}
		\hat{\beta}_t^*=
		\begin{cases} \hat{\beta}_t, &  \mbox{~if~} |\hat{\beta}_t| \le \log n,
			\\
			\mathrm{sign}(\hat{\beta}_t), & \mbox{~if~} |\hat{\beta}_t| > \log n,
		\end{cases}
	\end{equation}
	where
	\begin{equation}\label{hat-beta}
		\hat{\beta}_t=\frac{1}{2}\log\frac{ \sum_{i,j\neq t,i\neq j}A_{ti}B_{ij}A_{jt} }{ \sum_{i,j\neq t,i\neq j}B_{ti}A_{ij}B_{jt} }:=\frac{1}{2}\log\frac{T_{n,t}(a)}{T_{n,t}(b)},\quad t\in[n].
	\end{equation}
	In the above equation, we define
	\begin{equation}\label{definition-B}
		B=(B_{ij}):= J_n - A - \diag(J_n),
	\end{equation}
	where $J_n$ denotes the $n\times n$ matrix with all entries being equal to $1$.
	That is, $B$ records the information which nodes are not connected by edges.

	The estimator $\hat{\beta}_t$ in \eqref{hat-beta} is based on the log-ratio of $T_{n,t}(a)$ to $T_{n,t}(b)$.
	We explain why we use a threshold estimator in \eqref{hat-beta-star} to estimate $\beta_t$, rather than $\hat{\beta}_t$ itself.
	If the network is very sparse (e.g., there are only a few nodes having edges), then
	the value of $T_{n,t}(a)$ or $T_{n,t}(b)$ may be equal to zero for some $t\in [n]$.
	This will lead to an infinite number. A threshold estimator avoids such scenario to occur.
	When $\min_t |\beta_t| \ge \log n$,  the network density is either $O(1/n^2)$ or $1 - O(1/n^2)$,
	corresponding to a nearly empty graph with no edges or a nearly complete graph with all edges.
	Since such graphs obey far away from real-world networks, we use the
	the threshold $|\beta_t| <  \log(n)$.

	We now consider the computing cost of $\hat{\beta}_t$.
	According the definitions of $A$ and $B$ in \eqref{definition-B}, when $\{i, j, t\}$ are not distinct,
	$A_{ti}B_{ij}A_{jt}  =    B_{ti}A_{ij}B_{jt} = 0$. Therefore, we have the following lemma, whose proof is omitted.
	
	\begin{lemma}
		\label{lemma:compute}
		$T_{n,t}(a) =  (ABA)_{t,t}$ and $T_{n,t}(b) = (BAB)_{t,t}$.
	\end{lemma}
	
	The computation complexity of $\hat{\beta}_t$ depends mainly on computing
	$T_{n,t}(a)$ and $T_{n,t}(b)$, which are involved with algebraic products of three matrices.
	In sparse networks, the complexity of computing $T_{n,t}(a)$ and $T_{n,t}(b)$
	are $O(n d_{\max})$, where $d_{\max}$ is the maximal degree in the graph.
	For a very small $d_{\max}$, the complexity is approximately a linear order.

	\begin{remark}
		From Proposition \ref{propo:ab},
		we know that there exists
		many pairs $(a,b) \in S_m$ with $m \geq 3$ satisfying \eqref{critical}.
		As mentioned in the discussion section, we do not consider $m\ge 3$ due to three reasons.
		First,  it is much more complex to analyze the estimator for a larger $m\ge 3$,
		where it is involved with a very tediously long computation.
		Second, we have shown that  $\hat{\beta}_t^*$ is already rate optimal in next section.
		Third, numerical studies in Table \ref{table-compare-cycle} indicate there are no significant improvements on the estimation accuracy for the cases $m\ge 3$ against the case $m=2$.
	\end{remark}

	\section{Theoretical properties}
	\label{sec-theoretical}

	For convenience, define
	\begin{equation}\label{definition}
		\begin{array}{c}
			\theta=(\theta_1, \ldots, \theta_n)^\top = ( e^{\beta_1}, \ldots, e^{\beta_n} )^\top, \qquad
			\theta_{\max}=\max_i \theta_i, \qquad \theta_{\min}=\min_i \theta_i.
		\end{array}
	\end{equation}
	When estimating a single $\beta_t\in \{\beta_1, \ldots, \beta_n\}$, we assume
	\begin{align}\label{condition-main}
		\theta_{\max} \to 0,\qquad \theta_{t}\|\theta\|_1 \to \infty.
	\end{align}
	The first condition $\theta_{\max} \to 0$ guarantees that the network is sparse,
	which means that the network density goes to zero.
	This is in accord with most of observed networks and is also assumed in pervious works \cite[e.g.][]{Chen:2020,stein2025sparse,wang2025group}.
	The second condition guarantees good statistical properties of $\hat{\beta}_i$ including asymptotic optimality, consistency and asymptotic normality.
	When considering all $\beta_t$'s, $t\in [n]$, it could be replaced by the condition $\theta_{\min}\|\theta\|_1\to \infty$.

	\subsection{The signal-to-noise ratio of $\hat{\beta}_t^*$}
	\label{subsec:SNR}
	
	Recall $T_{n,t}(a)= \sum_{i,j\neq t,i\neq j}A_{ti}B_{ij}A_{jt}$ and $T_{n,t}(b)= \sum_{i,j\neq t,i\neq j}B_{ti}A_{ij}B_{jt}$.
	Define
	\begin{equation} \label{DefineUn}
		U_{n,t}(\beta)  =  T_{n,t}(a) -  e^{2\beta_t} T_{n,t}(b).
	\end{equation}
	Note that $\mathbb{E}[ T_{n,t}(a)] = e^{2\beta_t} \mathbb{E}[ T_{n,t}(b)]$. By basic calculus, we expect to see
	\[
	\hat{\beta}_t - \beta_t = \frac{1}{2}\log(\frac{ T_{n,t}(a) }{ e^{2\beta_t} T_{n,t}(b) } )    =   \frac{1}{2} \log(1 + \frac{U_{n,t}(\beta)}{ e^{ 2\beta_t } T_{n,t}(b)}) \approx
	\frac{U_{n,t}(\beta)}{2e^{2\beta_t} \mathbb{E}[ T_{n,t}(b)]}
	= \frac{ U_{n,t}(\beta) }{ 2\mathbb{E}[T_{n,t}(a)] }.
	\]
	Motivated by this, define the Signal-to-Noise Ratio (SNR) for $\hat{\beta}_t^*$ by
	\begin{equation} \label{DefineSNR}
		\mathrm{SNR} =  2\mathbb{E}[ T_{n,t}(a) ] / [\mathrm{Var}( U_{n,t}( \beta ) )]^{1/2},
	\end{equation}
	The variances of $T_{n,t}(a)$, $T_{n,t}(b)$ and $U_{n,t}( \beta )$ are stated in the follow theorem, whose proof
	is in Section A of the supplementary material.

	\begin{theorem}\label{theorem-variances}
		Suppose that condition \eqref{condition-main} holds. Then, for any $t\in[n]$, as $n\to \infty$, we have
		\begin{eqnarray}
			\label{thm-E-tna-tnb}
			\E[T_{n,t}(a)] & = & e^{2\beta_t} \E[T_{n,t}(b)] = ( 1 + o(1))\theta_t^2 \|\theta\|_1^2, \\
			\label{thm-var-tna}
			\mathrm{Var}(T_{n,t}(a)) & = & 4(1+o(1))\cdot\theta_t^3\|\theta\|_1^3,
			\\
			\label{thm-var-tnb}
			\mathrm{Var}(T_{n,t}(b)) & = & (1+o(1))\cdot(4\theta_t\|\theta\|_3^3+2)\|\theta\|_1^2,
			\\
			\label{thm-var-unt}
			\mathrm{Var}(U_{n,t}(\beta)) & = & (1+o(1))\cdot \mathrm{Var}(T_{n,t}(a)) =4(1+o(1))\cdot\theta_t^3\|\theta\|_1^3.
		\end{eqnarray}
	\end{theorem}

	By \eqref{thm-E-tna-tnb} and \eqref{thm-var-unt}, we have
	\[
	\mathrm{SNR}^2 = ( 1+ o(1))\cdot \frac{ 4\theta_t^4 \|\theta\|_1^4 }{ 4\theta_t^3\|\theta\|_1^3 } = ( 1+o(1)) \cdot (\theta_t \|\theta\|_1) = ( 1+o(1)) \cdot (1/r_n(\beta)),
	\]
	where we define
	\begin{equation}\label{definition-rn}
		r_n(\beta) = 1/( \theta_t \|\theta\|_1 ).
	\end{equation}
	When $\theta_t \|\theta\|_1 \to 0$,
	$\mathrm{SNR} \to 0$. In this case, the statistical properties of $\hat{\beta}_t$
	could not be guaranteed.
	Therefore, the assumption $\theta_t \|\theta\|_1 \to \infty$ in \eqref{condition-main} is neccessary.

	\subsection{Asymptotic optimality of $\hat{\beta}_t^*$}
	\label{subsec:est}
	In this section, we present the minimax risk of an estimator of $\beta_t$ and show that the CCR estimator achieves it.
	For a parameter space $\Theta(\epsilon_n)$, we define
	\[
	R_n^*(\epsilon_n) = \inf_{\hat{\beta}_t} \sup_{\beta \in \Theta(\epsilon_n) }
	\E [ ( \hat{\beta}_t - \beta_t)^2 ].
	\]
	
	We first study the upper bound of estimating $\beta_t$ by our approach in terms of the mean squared error (MSE).
	Recall that in condition \eqref{condition-main}, $\theta_t \|\theta\|_1 \to \infty$.
	We slightly strengthen this condition by assuming
	\begin{equation}
		\label{condition-upper-bound}
		\frac{ \theta_t \|\theta\|_1 }{ (\log n)^2 } \to \infty.
	\end{equation}
	The following theorem is about an upper bound of the MSE for $\hat{\beta}_t^*$, which is proved in Section B of the supplementary material.
	
	\begin{theorem} \label{thm:estUB}
		{\it (Upper bound)}.
		Suppose conditions \eqref{condition-main} and \eqref{condition-upper-bound} hold.
		As $n \to \infty$,
		\[
		\mathbb{E}[( \hat{\beta}_t^* - \beta_t )^2] \leq C \frac{\mathrm{Var}( U_{n,t}(\beta) ) }{ ( \mathbb{E}[ T_{n,t}(a) ] )^2} \leq   C r_n(\beta),
		\]
		where $r_n(\beta) \to 0$ is defined in \eqref{definition-rn}.
	\end{theorem}

	We now study the information lower bound by using the well-known two-point testing method  \cite[e.g.][]{DonohoLiu}.
	Specifically, we construct a null hypothesis $H_0: \beta=(\beta_1, \ldots, \beta_n)$ and an alternative hypothesis $H_1: \tilde{\beta}=(\beta_1, \ldots, \beta_{t-1}, \beta_t + \delta, \beta_{t+1}, \ldots, \beta_n)$.
	Here, the two parameters $\beta$ and $\tilde{\beta}$  differ by an amount of $\delta$ in only the $t$-th element.
	If the $\chi^2$-distance between the null distribution and alternative distribution goes to zero, then the minimax MSE is no smaller than $C \delta^2$ (which is then an information lower bound).
	Let $P_0^{(n)}$ and $P_1^{(n)}$ be the joint distribution of $A$ under the null $H_0$ and alternative $H_1$, respectively.
	Theorem \ref{theorem:estLB} is proved in Section C of the supplementary material.
	
	\begin{theorem}  \label{theorem:estLB}
		(Lower bound).  Suppose condition \eqref{condition-main}  holds and
		$\theta_t \|\theta\|_1 \delta ^2 \to 0$. Then, $\chi^2(P_0^{(n)}, P_1^{(n)}) \to 0$ as $n \to \infty$.
	\end{theorem}
	
	By \cite{DonohoLiu}, Theorem \ref{theorem:estLB} suggests a lower bound of $Cr_n(\theta)$ for the MSE,
	which matches the upper bound in Theorem \ref{thm:estUB}.
	Therefore, the lower bound is tight.
	This suggests that our method is asymptotic minimax in terms of the minimax framework for the MSE.
	In details, we consider a minimax setting as follows. Consider the parameter space
	\[
	\Theta_0 = \Big\{ \beta: \min\{ \frac{ 1 }{\theta_{\max}}, \frac{ \theta_t \|\theta\|_1 }{ (\log n)^2 } \} \ge \frac{1}{2} \log (\log n)  \Big\},
	\]
	where the term $\log(\log n)$ is chosen only for convenience and can be replaced by other diverging sequences.
	Note that if we neglect the $\log n$  terms in denominators on the left hand side, then $\Theta_0$
	defines essentially all $\beta$ satisfying condition \eqref{condition-main}. Fix an $\epsilon_n>0$ and
	consider the parameter space with the interesting range $n^{-2} \ll \epsilon_n \ll 1$ for $\epsilon_n$:
	\[
	\Theta( \epsilon_n ) = \{ \beta \in \Theta_0: r_n( \beta ) \le \epsilon_n \}.
	\]
	The minimax risk in $\Theta(\epsilon_n)$ is then $R_n^*(\epsilon_n) = \inf_{\hat{\beta}_t} \sup_{\beta \in \Theta(\epsilon_n) }
	\E [ ( \hat{\beta}_t - \beta_t)^2 ]$.
	Combining Theorem \ref{thm:estUB} and Theorem \ref{theorem:estLB}, we have the following result, whose proof is in Section D of the supplementary material.

	\begin{theorem} (Optimality)
		\label{thm-optimality}
		If $[\log (\log n) ]^2/n \le \epsilon_n \le 1 /\big\{ (\log n)^2 [\log (\log n)] \big\}$, then
		$C^{-1}\epsilon_n \le R_n^*(\epsilon_n) \le R_n(\hat{\beta}^*, \epsilon_n) \le C\epsilon_n$,
		where $R_n(\hat{\beta}_t^*; \Theta_{\epsilon_n}) =  \sup_{\beta \in \Theta(\epsilon_n)} \mathbb{E}[( \hat{\beta}_t^* - \beta_t )^2]$
	\end{theorem}

	\subsection{Asymptotic normality}
	Let
	\[
	V_{n,t}(\beta) = \sum_{i\neq t} P_{ti} \big( \sum_{j\neq i,t}Q_{ij}P_{jt} \big)^2. 
	\]
	By (25) in Lemma A.1 and (32) in Section A.1.1 of the supplementary material, we know
	\[
	V_{n,t}(\beta) = \mathrm{Var}(  \sum_{i,j\neq t,i\neq j}P_{ti}Q_{ij}W_{jt}  ) = (1 + o(1))\theta_t \|\theta\|_1^3.
	\]
	The asymptotic distribution of $\hat{\beta}_t$ is stated below, whose proof is in Section E of the supplementary material.
	
	\begin{theorem}
		\label{th-asy}
		Suppose that \eqref{condition-main} holds. For a fixed integer $k$, as $n\to\infty$,  we have
		\begin{align}\label{norm}
			\Big( \frac{ \hat{\beta}_{i_1}^* -\beta_{i_1} }{ \sigma_{i_1} }, \dots, \frac{ \hat{\beta}_{i_k}^* -\beta_{i_k}}{ \sigma_{i_k} } \Big)
			\rightsquigarrow  N(0,I_k),
		\end{align}
		where ``$\rightsquigarrow$" denotes ``convergence in distribution", $I_k$ is the $k\times k$ identity matrix,
		and $\sigma_t^2=V_{n,t}(\beta)/t_{n,t}^2(a)$.
	\end{theorem}

	\begin{remark}
		The convergence rate of $\hat{\beta}_t^*$ is $V_{n,t}^{1/2}(\beta)/t_{n,t}(a)$, where
		\[
		V_{n,t}(\beta)/t_{n,t}^2(a)=(1+o(1))/\sqrt{\theta_t \|\theta\|_1}.
		\]
		Therefore, the condition $\theta_t\|\theta\|_1\to \infty$ is necessary for guarantee the statistical properties of $\hat{\beta}_t$. Furthermore, according to \cite{Yan2013clt}, the convergence rate of the MLE of $\beta_t$ is
		$1/v_{ii}^{1/2}$, where $v_{ii} = \mathrm{Var}(d_i)$. If $\theta_{\max} \to 0$, then $v_{ii}= (1+o(1))\sum_j \theta_i \theta_j=(1+o(1))\theta_t \|\theta\|_1$, which is the same as the variance of $\hat{\beta}_t^*$. That's to say,
		the asymptotic variance of the CCR estimator is the same  as that of the MLE.
	\end{remark}

	Let
	\[
	\widehat{V}_{n,t}(\beta)=\sum_{i\neq t}A_{ti}\big(\sum_{j\neq i,t}B_{ij}A_{jt}\big)^2, \mbox{~~ and ~~ } \hat{\sigma}_t^2=\sum_{j=1}^k\frac{\widehat{V}_n(\beta_t)}{T_{n,t}^2(a)}
	\]
	\begin{theorem}
		\label{th-limit}
		If $\theta_{\max} \to 0$ and $\theta_t\|\theta\|_1/\log n \to \infty$, then
		\[
		\frac{ \widehat{V}_{n,t}(\beta) }{ V_{n,t}(\beta) } \stackrel{p}{\longrightarrow} 1,
		\qquad  \frac{ T_{n,t}(a) }{ t_{n,t}(a) } \stackrel{p}{\longrightarrow} 1.
		\]
		As a result, we have
		\begin{align}\label{norm}
			\Big( \frac{ \hat{\beta}_{i_1}^* -\beta_{i_1} }{ \hat{\sigma}_{i_1} }, \dots, \frac{ \hat{\beta}_{i_k}^* -\beta_{i_k}}{ \hat{\sigma}_{i_k} } \Big)
			\rightsquigarrow  N(0,I_k),
		\end{align}
		where $\hat{\sigma}_t^2 = \widehat{V}_{n,t}(\beta)/T_{n,t}^2(a)$.
	\end{theorem}

	The proof of the above theorem is in Section F of the supplementary material.
	It can be used to test a simple null $H_0: \beta_t = \beta_{t,0}$ or a homogeneous null $H_0: \beta_{i_1} = \cdots = \beta_{i_k}$ with a fixed $k$.
	We now give the power analysis for testing the simple null.
	Consider the hypothesis testing problem $H_0: \beta_t = \beta_{t,0} \leftrightarrow H_1: \beta_t = \beta_{t,1}$.
	We could construct the test statistic $\psi_t = (\hat{\beta}_t^*-\beta_{t,0})/\hat{\sigma}_t$.
	The power of the test statistic $\psi_t$ is approximate $2[1-\Phi(|\hat{\beta}_t^*-\beta_{t,0}|/\sigma_t]$, where
	$\Phi(x)$ denote the cumulative distribution function of the standard normality.
	Therefore, we have the following result, whose proof is in Section G of the supplementary material. 	
	
	\begin{theorem}(Power)\label{corollary-power}
		If $ \sigma_t^{-1}|\beta_{t,1} - \beta_{t,0}| \to 0$, then $\P( |\psi_t| \ge z_{1-\alpha/2}) \to \alpha$, i.e.,
		$\psi_t$ achives the nominal level, where $z_{1-\alpha/2}$ denotes the $(1-\alpha/2)$-quantile of the standard normal distribution.
		If $ \sigma_t^{-1}|\beta_{t,1} - \beta_{t,0}| \to \infty$, then $\P( |\psi_t| \ge z_{1-\alpha/2}) \to 1$, i.e., $\psi_t$ is asymptotically powerful.
	\end{theorem}

	\subsection{Uniform consistency}
	\label{subsec-uniform-con}
	In the above sections, we have studied the asymptotic properties of a single $\hat{\beta}_t^*$.
	We now move to discuss the uniform consistency for all the estimators $\hat{\beta}_t^*, t=1, \ldots, n$.
	To achieve it, we need a stronger condition than $\theta_t\|\theta\|_1\to\infty$ in \eqref{condition-main}.
	The uniform consistency result is stated as follows, whose proof is in Section H of the supplementary material.
	
	\begin{theorem}\label{th-consistency}
		Suppose that $\theta_{\max}\to 0$ and  $\theta_{\min}^2\|\theta\|_1\gg \log^{1/2} n$. Then, we have
		\begin{equation*}
			\max_{t}|\hat{\beta}_t^*-\beta_t| =O_p(c(\theta)\theta_{\min}^{-1/2}\|\theta\|_1^{-1/2}\log^{1/2} n   ) = o_p(1),
		\end{equation*}
		where $c(\theta)=\max\left\{1,
		\|\theta\|_2\|\theta\|_1^{-1}\theta_{\min}^{-1},
		\theta_{\min}^{-3/2}\|\theta\|_1^{-1/2} \right\}$.
	\end{theorem}

	\section{Numerical experiments}
	\label{section-simulation}
	
	In this section, we evaluate the performance of the proposed  estimator in networks of finite sizes.
	
	\subsection{Simulations}
	
	We consider the design for degree parameters:
	\[
	\beta_i \sim U(-1,1) + \gamma,\quad i=1, \ldots, n
	\]
	where the parameter $\gamma$ is used to control the sparsity level of networks and $U(-1,1)$ denotes the uniform distribution in the interval $[-1,1]$.
	Specifically, we chose three different values for the parameter $\gamma$:
	\[
	\gamma_1 = -\tfrac{1}{6} \log n, \quad \gamma_2=-\tfrac{1}{4}\log n + \tfrac{1}{7}\log(\log n), \quad
	\gamma_3 = - \tfrac{1}{2}\log n + \tfrac{1}{2}\log(\log n),
	\]
	covering from moderately sparse networks to extremely sparse networks.
	For $\gamma_1$, $\gamma_2$ and $\gamma_3$, the densities of generated networks are close to $1/n^{1/3}$, $1/n^{1/2}$ and $1/n$, respectively.
	When the network density is close to $1/n$, it implies that most of nodes have very small degrees.

	We first compare the CCR estimators based on subgraphs with $3$ nodes, $5$ nodes and $7$ nodes, where the estimator in the $3$-node-case is exactly given in
	\eqref{hat-beta-star}.
	The expressions of the estimators in the $5$-node-case and the $7$-node-case that we use, 
	are
	\[
	\hat{\beta}_{i_1}=\frac{1}{2}\log\frac{ \sum_{i_1,\dots,i_5(distinct)}A_{i_1i_2}B_{i_2i_3}A_{i_3i_4}B_{i_4i_5}A_{i_5i_1} }{ \sum_{i_1,\dots,i_5(distinct)}B_{i_1i_2}A_{i_2i_3}B_{i_3i_4}A_{i_4i_5}B_{i_5i_1} }, 
	\]
	and
	\[
	\hat{\beta}_{i_1}^\prime=\frac{1}{2}\log\frac{ \sum_{i_1,\dots,i_7(distinct)}A_{i_1i_2}B_{i_2i_3}A_{i_3i_4}B_{i_4i_5}A_{i_5i_6}B_{i_6i_7}A_{i_7i_1} }{ \sum_{i_1,\dots,i_7(distinct)}B_{i_1i_2}A_{i_2i_3}B_{i_3i_4}A_{i_4i_5}B_{i_5i_6}A_{i_6i_7}B_{i_7i_1} }, 
	\]
	respectively.
	We consider three sizes for the number of nodes $n=1000, 2000, 3000$.
	Each simulation is repeated $500$ times.
	The simulation results are reported in Table \ref{table-compare-cycle}.
	From this table, we can see: (1)
	When $n=1000$, the estimator in the $3$-node-case has a little larger error than the other two estimators. (2) The errors of the three different estimators
	are comparable for $n=2000$ and $n=3000$.
	This shows that it does not bring the major improvements when using larger $m$-cycle counting approaches in case of large sizes of networks.

	\begin{table}[!htpb]
		\centering
		\caption{The reported values are the average absolute errors of the CCR estimators $\hat{\beta}_i^*$,
			$\hat{\beta}_i^{(5)}$ (subgraphs with $5$ nodes) and $\hat{\beta}_i^{(7)}$ (subgraphs with $7$ nodes), based on $100$ generated networks in each simulation.
			$\|\cdot\|_\infty$ means $\|\hat{\beta}-\beta\|_\infty$. }
		\label{table-compare-cycle}
		\renewcommand\arraystretch{1.3}
		\scalebox{0.75}{
			\begin{tabular}{l lll  l lll l lll l lll}
				\hline
				$\gamma$ & \multicolumn{3}{c}{$i=1$} & & \multicolumn{3}{c}{$i=n/2$} & &
				\multicolumn{3}{c}{$i=n$}  & & \multicolumn{3}{c}{$\|\cdot\|_\infty$}
				\\
				\cline{2-4} \cline{6-8}  \cline{10-12} \cline{14-16}
				&  $\hat{\beta}_i^*$  &  $\hat{\beta}_i^{(5)}$  &   $\hat{\beta}_i^{(7)}$ &
				&
				$\hat{\beta}_i^*$  &  $\hat{\beta}_i^{(5)}$  &   $\hat{\beta}_i^{(7)}$ &
				&
				$\hat{\beta}_i^*$  &  $\hat{\beta}_i^{(5)}$  &   $\hat{\beta}_i^{(7)}$ &
				&
				$\hat{\beta}_i^*$  &  $\hat{\beta}_i^{(5)}$  &   $\hat{\beta}_i^{(7)}$
				\\
				\hline
				& \multicolumn{15}{c}{n=1000}
				\\
				\cline{2-4} \cline{6-8}  \cline{10-12} \cline{14-16}
				$\gamma_1$ & $0.086 $&$  0.085$&$ 0.085$ &
				&
				$0.108$ & $0.108$ & $0.108$  &
				&
				$0.1$ & $0.099$ & $0.099$    &
				&
				$0.508$ & $0.491$ & $0.491$
				\\
				$\gamma_2$ & $0.092$&$0.091$&$0.091$ &
				&
				$0.1$ & $0.1$ & $0.1$  &
				&
				$0.093$ & $0.093$ & $0.093$          &
				&
				$0.726$ & $0.685$ & $0.685$
				\\
				$\gamma_3$ & $0.34$&$0.307$&$0.307$ &
				&
				$0.43$ & $0.401$ & $0.401$ &
				&
				$0.435$ & $0.404$ & $0.404$       &
				&
				$1.795$ & $1.967$ & $1.968$
				\\
				\hline
				& \multicolumn{15}{c}{n=2000}
				\\
				\cline{2-4} \cline{6-8}  \cline{10-12} \cline{14-16}
				$\gamma_1$ & $0.063$&$0.063$&$0.063$ &
				&
				$0.079$ &$ 0.078$ &$0.078$  &
				&
				$0.073$ &$0.073$ &$0.073$   &
				&
				$0.412$ &$0.404$ &$0.404$
				\\
				$\gamma_2$ & $0.088$&$0.087$&$0.087$ &
				&
				$0.118$ &$0.116$ &$0.116$  &
				&
				$0.109$ &$0.108$ &$0.108$  &
				&
				$0.643$ &$0.616$ &$0.616$
				\\
				$\gamma_3$ & $0.347$&$0.313$&$0.313$ &
				&
				$0.412$ &$0.375$ &$0.375$ &
				&
				$0.351$ &$0.3$ &$0.3$  &
				&
				$1.912$ & $2.052$ &  $2.053$
				\\
				\hline
				& \multicolumn{15}{c}{n=3000}
				\\
				\cline{2-4} \cline{6-8}  \cline{10-12} \cline{14-16}
				$\gamma_1$ & $0.056$&$0.055$&$0.055$ &
				&
				$0.044$ & $0.044$ & $0.044$  &
				&
				$0.07$ & $0.069$ & $0.069$      &
				&
				$0.39$ & $0.381$ & $0.381$
				\\
				$\gamma_2$ & $0.082$&$0.081$&$0.081$ &
				&
				$0.064$ & $0.063$ & $0.063$  &
				&
				$0.105$ & $0.103$ & $0.103$     &
				&
				$0.603$ & $0.578$ & $0.578$
				\\
				$\gamma_3$ & $0.285$&$0.271$&$0.271$ &
				&
				$0.201$ & $0.192$ & $0.192$ &
				&
				$0.339$ & $0.389$ & $0.389$      &
				&
				$1.934$ & $2.112$ & $2.113$
				\\
				\hline
			\end{tabular}
		}
	\end{table}

	Second, we evaluate the estimation error of the estimator $\hat{\beta}_i^*$ and compare it with the MLE, denoted by $\hat{\beta}^{mle}_i$.
	We do not compare it with the $L_0$-penalized MLE in the sparse $\beta$-model \citep{Chen:2020} or regularized MLE \citep{shao2023l2} since
	the former is computationally time-consuming and depends on a sparsity structure assumption on the unknown parameters, and the latter incurs the bias problem as demonstrated in the simulations of  \cite{shao2023l2}.
	We consider three choices for the network size $n$: $n=500, 1000, 2000$.
	Each simulation was repeated $100$ times. 
	We record the average values of the absolute errors for the CCR estimator and the MLE,
	i.e., $|\hat{\beta}_i^*-\beta_i|$ and $|\hat{\beta}_i^{mle} - \beta_i|$, respectively.
	Further, we compare the average $L_1$-norm and $L_\infty$-norm error for $\hat{\beta}_i^*$ and $\hat{\beta}_i^{mle}$.
	We also record the average network density, i.e., $\sum_{i\neq j}A_{ij}/(n(n-1))$, and the frequencies that the MLE fails to exist.
	Remark that when there exists at least one degree $d_i$ being equal to $0$, the MLE does not exist.
	Therefore, the values of $\hat{\beta}_i^{mle}$ are computed conditional on the event that the MLE exists.
	The simulation results are shown in Table \ref{table-compare-A}.

	\begin{table}[!htpb]\centering
		\caption{
			The reported values are the average value of $|\hat{\beta}_i-\beta_i|$, the average value of $|\hat{\beta}^{mle}_i-\beta_i|$,
			the averagve network density ("Density") and
			the frequency that the MLE fails to exist ("Fail-freq").
			$\|\infty\|$ means $\|\hat{\beta}-\beta\|_\infty$ or $\|\hat{\beta}^{mle} - \beta \|_\infty$.
			$\|\cdot\|_{(1)}$ means $\|\hat{\beta}-\beta\|_1/n$ or $\|\hat{\beta}^{mle} - \beta \|_1/n$.
			``NA" means that the MLE failed to exist in all generated networks.
		}
		\label{table-compare-A}
		\scriptsize
		\renewcommand\arraystretch{1.3}
		\scalebox{0.92}{
			\begin{tabular}{l l ll l l ll lll ll}
				\hline
				$\gamma$   & \multicolumn{2}{c}{$i=1$}  &  \multicolumn{2}{c}{$i=n/2$}
				& \multicolumn{2}{c}{$i=n$}    &   \multicolumn{2}{c}{$\|\cdot\|_{(1)}$}  &
				\multicolumn{2}{c}{$\|\cdot\|_\infty$}  & Density & Fail-freq
				\\
				\cline{2-3}  \cline{4-5} \cline{6-7} \cline{8-9}  \cline{10-11}
				\\
				& CCR & MLE & CCR & MLE & CCR & MLE & CCR & MLE & CCR & MLE & &
				\\
				\cline{2-13}
				& \multicolumn{12}{c}{$n=500$}      \\
				\cline{2-13}
				$\gamma_1$   & $0.093 $&$ 0.094 $  & $0.125  $&$ 0.126 $ & $0.096 $&$ 0.096 $
				& $0.115 $ & $ 0.115 $ & $0.59 $ & $0.59$ & $0.131$ & $0$
				\\
				$\gamma_2$  & $0.188 $ & $ 0.19$  & $0.158$ & $0.157$ & $0.158 $ & $ 0.159 $
				& $0.142 $ & $ 0.142 $ & $0.818$ & $0.817$ & $0.092$ & $0$
				\\
				$\gamma_3$    & $0.275$ & NA  & $0.194$ & NA  & $0.463$  &  NA  & $0.324$ &  NA
				& $1.775$ & NA  & $0.016$ & $0.97$
				\\
				\cline{2-13}
				&\multicolumn{12}{c}{$n=1000$}\\
				\cline{2-13}
				$\gamma_1$   & $0.102 $&$ 0.101$ & $0.09 $&$ 0.091 $
				& $0.107 $&$ 0.106 $ & $0.088 $&$ 0.088 $ & $0.495$&$0.495$ & $0.108$ & $0$
				\\
				$\gamma_2$  & $0.181 $&$ 0.182 $ & $0.185 $&$ 0.185 $
				& $0.126 $&$ 0.126$  & $0.117 $ &$ 0.117 $ & $0.731$ & $0.731$ & $0.069$ & $0$
				\\
				$\gamma_3$    & $0.255 $& NA  & $0.361$ & NA  & $0.429$ &  NA
				& $0.31$ & NA  & $1.902$ & NA & $0.009$ & $1$
				\\
				\cline{2-13}
				&\multicolumn{12}{c}{$n=2000$}\\
				\cline{2-13}
				$\gamma_1$   & $0.054 $&$ 0.053 $ & $0.06 $&$ 0.06 $ & $0.071 $&$ 0.071 $
				& $0.069 $ & $ 0.069 $ & $0.407$ & $0.406$ & $0.091$ & $0$
				\\
				$\gamma_2$  & $0.11 $ & $ 0.109 $ & $0.079 $ & $ 0.079 $ & $0.075$ & $ 0.075$
				& $0.091 $ & $ 0.091 $ & $0.586$ & $0.586$ & $0.051$ & $0$
				\\
				$\gamma_3$    & $0.302$ & NA  & $0.274 $& NA  & $0.464 $& NA
				& $0.304 $& NA  & $1.987$& NA & $0.005$ & $1$
				\\
				\hline
			\end{tabular}
		}
	\end{table}

	\begin{table}[!htpb]
		\centering
		\caption{
			The reported values are the $95\%$ coverage frequencies of $\beta_i-\beta_j$. }
		\label{table-coverage}
		\renewcommand\arraystretch{1.3}
		\begin{threeparttable}
			\begin{tabular}{ll ll l}
				\hline
				$n$   &  $(i,j)$   & $\gamma_1$  & $\gamma_2$     &  $\gamma_3 $                                 \\
				\cline{3-5}
				$500$    &  $(1,2)$   & 0.93 & 0.94 & 0.945
				\\
				& $(\tfrac{n}{2}-1, n/2)$  & 0.902 & 0.94 & 0.948
				\\
				& $(n-1,n)$    & 0.913 & 0.934 & 0.944
				\\
				$1000$    &  $(1,2)$   & 0.939 & 0.945 & 0.944
				\\
				&$(\tfrac{n}{2}-1, n/2)$  & 0.925 & 0.953 & 0.934
				\\
				& $(n-1,n)$   & 0.929 & 0.944 & 0.941
				\\
				$2000$    &  $(1,2)$   & 0.937 & 0.953 & 0.946
				\\
				& $(\tfrac{n}{2}-1, n/2)$  & 0.937 & 0.941 & 0.943
				\\
				& $(n-1,n)$    & 0.945 & 0.941 & 0.971
				\\
				\hline
			\end{tabular}
		\end{threeparttable}
	\end{table}

	\begin{figure}[!htbp]
		\centering
		\includegraphics[height = 4in]{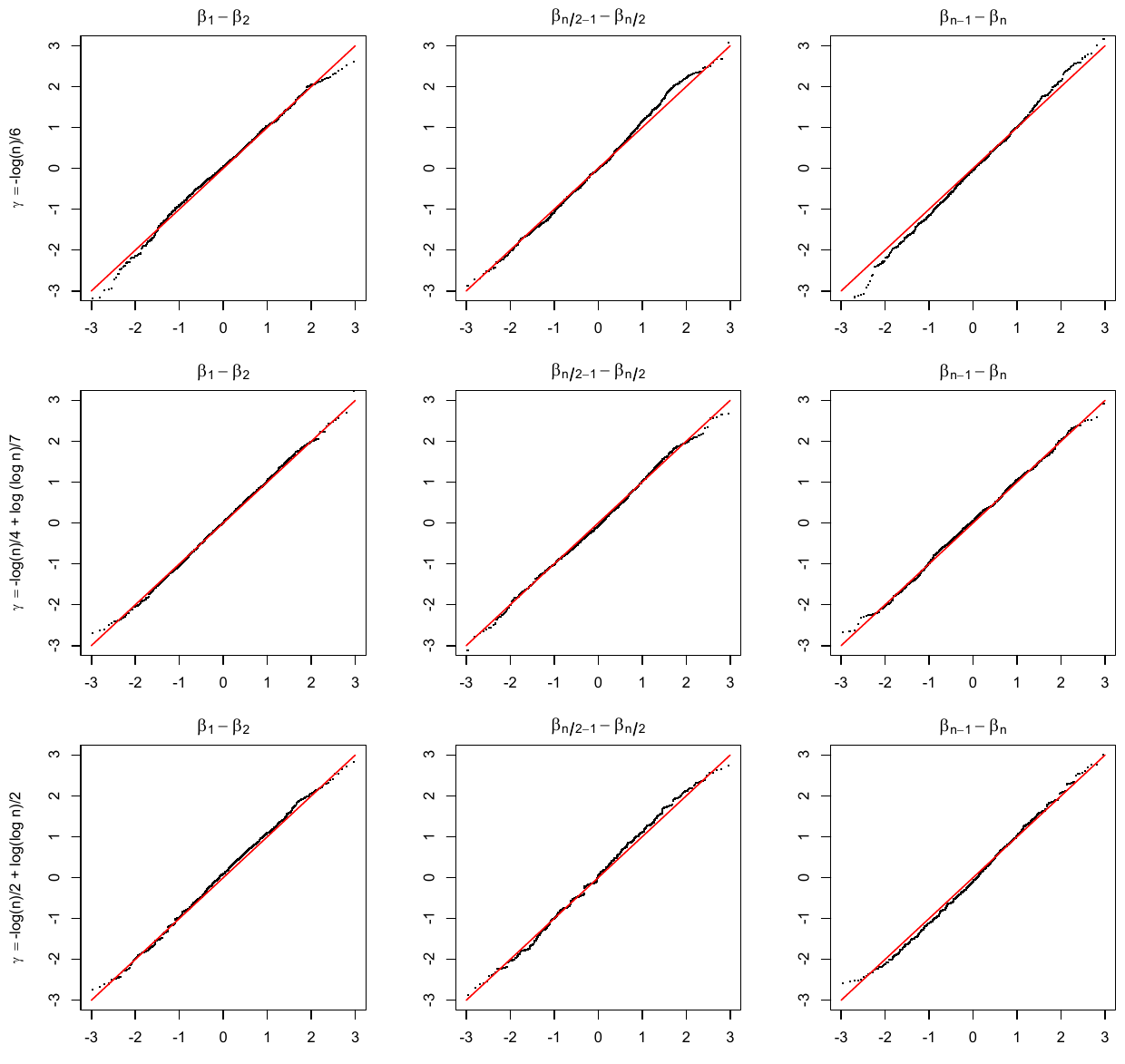}
		\caption{QQ-plots of $(\hat{\beta}_i-\hat{\beta}_j-(\beta_i-\beta_j))/\hat{\sigma}_{ij}$ ($n=1000$).}
		\label{fig-norm}
	\end{figure}

	From Table \ref{table-compare-A},
	we can see that all the estimation errors between the CCR estimator and the MLE
	are very close when $\gamma_1=-\tfrac{1}{6}\log n$ or $\gamma_2=-\tfrac{1}{4}\log n + \tfrac{1}{7}\log (\log n)$. In addition,
	there is a notable decreasing trend for the errors when $n$ increases from $500$ to $2000$ and
	$\gamma$ is fixed.
	When $n$ is fixed, the estimation become less accuracy when the network becomes more sparse.
	On the other hand,
	when $\gamma_3=-\tfrac{1}{2}\log n + \tfrac{1}{2}\log (\log n)$,
	implying a extremely sparse networks (the network density less than $1\%$),
	the MLE failed to exist in nearly all generated network data while the CCR estimator
	deviate from the true parameter but not too much.
	These observations indicates that our proposed estimator performs well
	and is comparable to the MLE in  sparse networks.

	Next, we evaluate the asymptotic normality.
	We draw the quantile-quantile (QQ) plots for the pairwise difference of the CCR estimator
	$\hat{\beta}_i-\hat{\beta}_j$. According to Theorem \ref{th-limit}, we know
	\begin{equation}
		\label{eq-test}
		\hat{\xi}_{ij}=\frac{ (\hat{\beta}_i-\hat{\beta}_j) - (\beta_i-\beta_j) }{ \sqrt{\hat{\sigma}_{i}^2+\hat{\sigma}_{j}^2} } \rightsquigarrow N(0,1),
	\end{equation}
	where $\hat{\sigma}_{i}^2$ is defined in \eqref{norm}.
	We show the QQ plots in case of $n=1000$ in Figure \ref{fig-norm}.
	The other plots in cases of $n=500$ or $n=2000$ are similar and not shown to save space.
	From  Figure \ref{fig-norm}, we can see that the empirical quantiles of $\hat{\xi}_{ij}$
	agree with the quantiles of the standard normality well.
	Furthermore, we record the $95\%$ coverage frequencies of $(\beta_i-\beta_j)$ in Table \ref{table-coverage}. 
	From this table, we can see that all frequencies are close to $0.95$, except that several values
	deviate from the target level a little.

	\subsection{A real data analysis}
	
	\cite{Mcauley2012Facebook} collected an undirected ego network data amongst users in a Facebook application,
	which is available at  \url{http://snap.stanford.edu/data/ego-Facebook.html}.
	We use this data set to illustrate the practical application of out proposed method.
	This data set has $4,039$ nodes denoting users in the Facebook application  and $88,234$ edges recording friendships between users,
	where the 'ego' node is removed.
	This is a very sparse network because the network density equals $88234/(4039\times 4038/2)\approx 0.011$.
	The maximum, $3/4$-quantile, the median and $1/4$-quantile, the minimum of the degrees
	are $1045,57,25,11,1$, respectively.

\begin{figure}[!htbp]
		\centering
		\includegraphics[height = 2.5in]{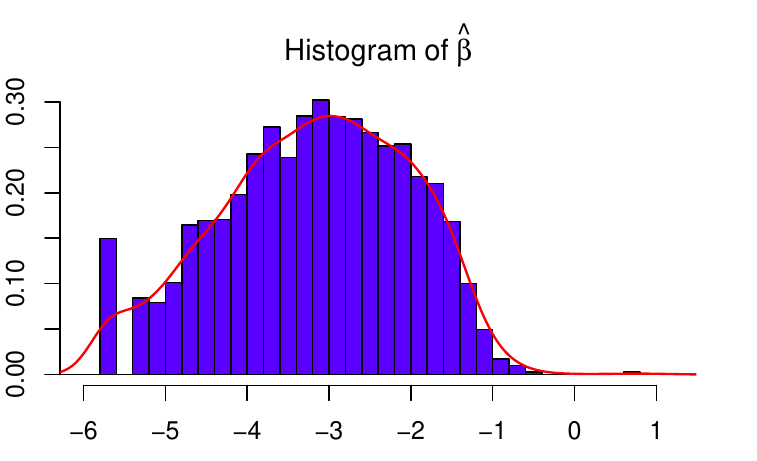}
		\caption{Histogram of the $\hat{\beta}$, with
			the red color denoting the density estimator.}
		\label{fig-facebook}
	\end{figure}

	Recall the CCR estimator in \eqref{hat-beta}.
	It is  based on the logarithm of the ratio between two counting numbers $T_{n,t}(a)$ and $T_{n,t}(b)$.
	When any counting number is equal to zero, it shows that the corresponding node is either isolated or has few edges such that
	its degree parameter is very negatively large and could not be accurately estimated.
	In this case, the leading estimator is meaningfulness.
	So we only consider those nodes, where the values of $T_{n,t}(a)$ and $T_{n,t}(b)$ are both not equal to zeros.
	There are $3,674$ such nodes in the data set. We calculated the estimators of these nodes and draw their
	histogram in Figure \ref{fig-facebook}.
	This histogram indicates an irregular distribution for the parameters and there are a few nodes
	having large negative values.
	The minimum, $1/4$-quantile, median, $3/4$-quantile and maximum values for $\hat{\beta}_t$ are $-8.304, -4.302, -3.2664, -2.362$ and $1.1$.
	
	Next, we test the equality of several pairs of parameters.
	We use the test statistic in \eqref{eq-test} to test the null
	$H_{0}: \beta_{i}=\beta_{j}$ against the alternative $H_{1}: \beta_{i}\neq\beta_{j}$.
	At level $\alpha=0.05$, the hypothesis can be rejected if
	$|\hat{\xi}_{ij}|>z_{1-\alpha/2}$, where $z_\tau$ denotes the $\tau$-quantile of
	the standard normality.
	We random select $10$ nodes from the set $\{1,2,\dots,1848\}$ and other $10$ nodes from the set $\{1849,\dots,3697\}$.
	Then, we got ten pairs $(937,3967)$, $(1867,3934)$, $(920,3034)$, $(1877,2486)$, $(992,2255)$, $(60,2941)$, $(1807,2437)$, $(1450,2728)$, $(1322,2948)$ and $(9,2301)$.
	The results for testing the homogeneity of these pairwise parameters
	are showed in Table \ref{tab-power}.
	From this table, we can see that the differences of the degree parameters between pairs $(937, 3697)$, $(1867, 3934)$, $(1877, 2486)$ and $(1322, 2948)$ are not significant, while those for other pairs
	are significant.
	
	\begin{table}
		\caption{Result for the multiple hypothesis test. The first four columns show the comparison with $V_1$ (left) and $V_2$ (right).}
		\renewcommand\arraystretch{1.3}
		\centering
		\begin{tabular}{c c c c c}\label{tab-power}
			Node	& Degree & $\hat{\beta}_t$ & $\hat{\sigma}_{\beta_t}^2$ & p-value  \\
			\hline
			(937, 3697) & (37, 15) & (-2.935, -4.336) & (0.142, 0.649)  & 0.0578\\
			(1867, 3934) &  (31, 22) & (-2.936, -3.402) &  (0.152, 0.244)  & 0.229 \\
			(920, 3034) &  (29, 88) & (-3.674, -1.932) & (0.353,   0.054)  & 0.003 \\
			(1877, 2486) &  (56, 137) &  (-2.26, -1.913) &  (0.08,   0.049) & 0.166 \\
			(992, 2255)  &   (19, 90) &  (-3.31, -2.014) & (0.228,   0.057) & 0.008\\
			(60, 2941) & (19, 64) & (-3.87, -2.326) & (0.393,  0.076) & 0.012\\
			(1807, 2437) & (32, 94) & (-2.88,  -2.044) & (0.144,  0.055)  & 0.031 \\
			(1450, 2728)  &  (90, 13) & (-1.91, -3.784) &  (0.055, 0.38) & 0.002\\
			(1322, 2948) & (25, 17) & (-3.15, -3.893) & (0.19, 0.364) & 0.159 \\
			(9, 2301)  & (8, 122)  &  (-4.593, -1.688) &  (0.815,   0.046) & 0.001 \\
			\hline
		\end{tabular}
	\end{table}
	
	\section{Discussion}\label{section-discussion}	
	
	We have proposed a cycle counting approach to estimate all the unknown parameters in the $\beta$-model.
	The CCR estimator has an explicit expression in terms of the logarithmic ratio of two counts of two subgraphs,
	where the counts are simple algebraic operations. Therefore, it can be scale to large networks.
	In theory, we have established its asymptotic optimality, where the CCR estimator achieves the minimax risk in terms of the mean
	square error. We have also derived the
	uniform consistency for all the estimators of parameters and asymptotic normality
	for a fixed number of estimators under mild conditions in sparse networks.

	Although we establish the central limit theorem for the estimators $(\hat{\beta}_{i_1}, \ldots, \hat{\beta}_{i_k})$, they are finite dimensional.
	It is nontrivial to extend it to an increasing dimensional case.
	In the fixed dimension $k$, the estimators $\hat{\beta}_{i_1}, \ldots, \hat{\beta}_{i_k}$
	can be represented as $k$ independent sums of random variables with vanishing remainder terms; see (100) in the supplementary material.
	If $k$ increases with a slow rate, the subgraph counting statistics in the expressions of $(\hat{\beta}_{i_1}, \ldots, \hat{\beta}_{i_k})$
	share minor same terms. This will have weak influence on the covariances among the estimators.
	It is expected that the estimators $\hat{\beta}_{i_1}, \ldots, \hat{\beta}_{i_k}$ are still asymptotically independent.
	If $k$ quickly increases, then it is unclear whether the correlations between different estimators could be neglected or not.
	It is of interest to investigate this issue.
	Another interesting issue is to see whether the cycle counting approach could be used to estimate the parameters in a general
	exponential random graph model.

	\setlength{\itemsep}{-1.5pt}
	\setlength{\bibsep}{0ex}
	\bibliographystyle{apalike}
	\bibliography{ref}

\end{document}